\documentclass{emulateapj}
\usepackage{amssymb,amsmath,amstext,amsgen,amsopn,amsxtra,indentfirst,lscape}
\shorttitle{XMM-NEWTON 87A}
\shortauthors{Heng et al.}
\begin{document}

\title{PROBING ELEMENTAL ABUNDANCES IN SNR 1987A USING XMM-NEWTON}

\author{Kevin Heng\altaffilmark{1,2}, Frank Haberl\altaffilmark{1},
Bernd Aschenbach\altaffilmark{1} \& G\"{u}nther Hasinger\altaffilmark{1}}

\altaffiltext{1}{Max Planck Institut f\"{u}r extraterrestrische Physik, Giessenbachstra$\beta$e, 85478 Garching, Germany}

\altaffiltext{2}{Max Planck Institut f\"{u}r Astrophysik, Karl-Schwarzschild-Stra$\beta$e 1, 85740 Garching, Germany}

\begin{abstract}
We report on the latest (2007 Jan) observations of supernova remnant (SNR) 1987A
from the {\it XMM-Newton} mission.  Since the 2003
May observations of Haberl et al. (2006), 11 emission lines have
experienced increases in flux by factors $\sim 3$ to 10 ($6 \pm 0.6$ on average), with the 775 eV line of O~{\sc
viii} showing the greatest increase.  Overall, we are able to
make Gaussian fits to 17 emission lines in the {\it RGS} spectra and
obtain line fluxes; we have observed
6 lines of Fe~{\sc xvii} and Fe~{\sc xviii} previously unreported by
{\it XMM-Newton}.  A two-shock model representing
plasmas in non-equilibrium ionization is fitted to the {\it EPIC-pn} spectra,
yielding temperatures of $\sim 0.4$ and $\sim 3$ keV, as well as
elemental abundances for N, O, Ne, Mg, Si, S and Fe.  

We demonstrate that the abundance ratio
of N and O can be constrained to $\lesssim 20\%$ accuracy $\left(\mbox{N/O} =
1.17 \pm 0.20 \right)$.  Within the same confidence interval, the same analysis suggests that the C+N+O
abundance varies from $\sim 1.1$ to $1.4 \times 10^{-4}$, verifying the {\it Chandra} finding by Zhekov et
al. (2006) that the C+N+O abundance is lower by a factor $\sim 2$
compared to the value obtained in the optical/ultraviolet study by
Lundqvist \& Fransson (1996).  Normalizing our
obtained abundances by the Large Magellanic Cloud (LMC) values of Hughes, Hayashi \& Koyama (1998), we find that O, Ne, Mg and Fe are under-abundant, while Si and S are over-abundant, consistent with the findings of Aschenbach (2007).  Such a result has implications for both the single-star and binary accretion/merger models for the progenitor of SNR 1987A.  In the context of the binary merger scenario proposed by Morris \& Podsiadlowski (2006, 2007), material forming the inner, equatorial ring was expelled after the merger, implying that either our derived Fe abundance is inconsistent with typical LMC values or that iron is under-abundant at the site of Sanduleak -69$^\circ$202.
\end{abstract}


\section{INTRODUCTION}
\label{sect:intro}

SNR 1987A is the Rosetta Stone (Allen 1960) of Type II supernova
remnants, resolved and well-studied in multiple wavebands, including the infrared
(Bouchet et al. 2006; Kjaer et al. 2007), optical/ultraviolet
(Gr\"{o}ningsson et al. 2007; Heng 2007) and radio (Gaensler et
al. 2007).  The physical mechanism
partially powering
optical/ultraviolet emission from the reverse shock is the same as the
one at work in Balmer-dominated supernova remnants (Heng \& McCray
2007; Heng et al. 2007).  The detection of a neutrino burst confirmed the core
collapse nature of the progenitor (Koshiba et al. 1987; Svoboda et al.
1987), though a
pulsar has yet to be detected (Manchester 2007).  A system of three
rings may be the result of a binary merger between two
massive stars about 20,000 years prior to the supernova explosion
(Morris \& Podsiadlowski 2006, 2007; hereafter MP0607).  Reviews of the multi-wavelength studies of SNR 1987A can be found in McCray (1993, 2005, 2007).

Mixing of the stellar envelope and core by Rayleigh-Taylor instabilities within the progenitor star, Sanduleak -69$^\circ$202, has been invoked to explain the early emergence of the 847 keV $\gamma$-ray line from SNR 1987A,
which was predicted by Shibazaki \& Ebisuzaki (1988) to reach its peak
around 1.1 years after the explosion, if one assumes a mixed mass of
about 5$M_\sun$ (Ebisuzaki \& Shibazaki 1988).  Instead, Matz et al. (1988)
observed the 847 keV line $\sim$ 6 to 8 months
post-explosion, suggesting even more extensive mixing of $^{56}$Co
than assumed.  A similar explanation (Ebisuzaki \& Shibazaki 1988) was given for the
early emergence of 16 to 28 keV X-rays (Sunyaev et al. 1990; Inoue et al. 1991).  The
$\gamma$-rays originate from the radioactive decay of $^{56}$Co, while the X-rays are from the Compton
degradation of the $\gamma$-rays (McCray, Shull \& Sutherland 1987).  

In the soft X-ray, SNR 1987A was first observed by Beuermann, Brandt \&
Pietsch (1994).  Subsequently, Hasinger, Aschenbach \&
Tr\"{u}mper (1996) tracked a steady increase of the soft X-ray
flux over 4 years with {\it ROSAT}.  Extensive work has since been
done by the {\it Chandra} (Michael et al. 2002; Park et al. 2002, 2004, 2005 [hereafter
P05], 2006, 2007 [hereafter P07]; Zhekov et al. 2005, 2006 [hereafter Z06]) and {\it XMM-Newton} groups
(Haberl et al. 2006, hereafter H06; Aschenbach 2007, hereafter A07).
The general picture gleaned from these studies is of a bimodal plasma
distribution present in the region between the forward and reverse
shocks (Fig. \ref{fig:87a}).  The soft X-rays ($\sim$ 0.3 to 0.5 keV) are from the
decelerated shock front interacting with dense protrusions
(``fingers'') on the
inner, equatorial ring, while the ``hard'' X-rays ($\sim$ 2 to 3 keV) are from a fast shock
propagating into more tenuous material.  The soft X-rays appear to be
correlated with optical ``hot spots'', believed to be emission from the shocked fingers, appearing around the equatorial
ring, while the hard X-ray and radio images exhibit structures that
coincide.  Between Days 6000 and 6200, the soft X-ray light curve
experienced an upturn and departure from an exponentially increasing
profile, which P05 interpreted as evidence that the blast wave had
reached the main body of the dense circumstellar material of the
equatorial ring.  Future studies on the nature of the soft X-ray light
curve are relevant to the issue of pre-ionization of the supernova
ejecta, which can potentially extinguish H$\alpha$ and Ly$\alpha$ emission from the reverse
shock (Smith et al. 2005; Heng et al. 2006).

In the studies described, relatively little attention has been paid to
the subject of inferring elemental abundances from X-ray analyses.
Fits to the X-ray spectra yield N, O, Ne,
Mg, S, Si and Fe abundances; such results have been tabulated and
studied by Z06 and H06, using {\it Chandra} and {\it XMM},
respectively.  Here, we examine such an
approach using the latest {\it XMM} data set of SNR 1987A, taken in
early 2007.  In \S\ref{sect:data}, we describe our observations and
data reduction techniques.  Our results are presented in
\S\ref{sect:results}.  In \S\ref{sect:discussion}, we perform a
detailed error analysis of individual abundances and their ratios.
We find that the N/O ratio as well as the individual N
and O abundances can be constrained to $\sim 20\%$ accuracy.  Our derived elemental abundances for O, Ne, Mg and Fe are under-abundant, while Si and S are over-abundant, relative to typical values for the Large Magellanic Cloud (LMC; Hughes, Hayashi \& Koyama 1998).  Such a result has implications for modeling the progenitor of SNR 1987A, which we discuss.

\section{OBSERVATIONS \& DATA REDUCTION}
\label{sect:data}

The latest XMM-Newton (Jansen, Lumb \& Altieri 2001) observations of SNR 1987A were performed
from 2007 Jan 17 18:23 to Jan 19 01:18 UT.   For our spectral
analysis, we utilize data from the CCD cameras {\it EPIC-pn} (Str\"uder et al. 2001a,b)
and {\it EPIC-MOS} (Turner et al. 2001), and the {\it Reflection
Grating Spectrometers} ({\it RGS}; den Herder et al. 2001).
Additional details and net exposure times for the different instruments are summarized in Table \ref{tab:obs}.

The data were processed with the {\it XMM-Newton Science Analysis
Software} ({\tt SAS}) version 7.1.0.  Using {\tt ds9}, we extracted source and
background {\it EPIC} spectra of SNR 1987A from circular regions placed on
the source and a proximate, point source-free area
(Fig. \ref{fig:xmm87a}).  For the {\it EPIC-pn} spectra, single-pixel
(PATTERN 0) events were selected; for {\it EPIC-MOS}
all valid event patterns (PATTERN 0 to 12) were used.  {\it RGS} spectra were produced using 
{\tt rgsproc}.  The spectra were binned to contain a minimum of 20 and 30 counts per
bin for {\it EPIC} and {\it RGS}, respectively.

\section{RESULTS}
\label{sect:results}

\subsection{SPECTRAL FITTING}
\label{subsect:spectral}

Spectral fitting was done using {\tt XSPEC} (Arnaud 1996) version 11.3.2.  We fit the reduced spectra using models with two components: a low ($T_{\rm{low}}$) and a high ($T_{\rm{high}}$) temperature
component.  Two models are considered in {\tt XSPEC}: {\tt
VNEI+VRAYMOND} and {\tt VPSHOCK+VPSHOCK}.  In H06, the former (``Model
A'') is seen
to give an excellent fit (``reduced chi-square'' of $\chi^2_r \approx
1.1$) to the {\it EPIC-pn} data.  However, the {\tt VRAYMOND}
component of the model, which describes a plasma in ionization
equilibrium, is at odds with the belief that both plasmas are in
non-equilibrium ionization (NEI; Z06); we note that Park et
al. (2004, 2006) found the low-temperature plasma component to be in a
highly-advanced ionization state (with ionization ages $\sim 10^{12}$
to $10^{13}$
cm$^{-3}$ s) and can be described by a thermal plasma in collisional
ionization equilibrium.  The {\tt VNEI} model
is a somewhat unsatisfactory representation of the physical situation
in the post-shock plasma of SNR 1987A, because it considers
only one value of the ionization age/time, $\tau$, for the entire plasma.  The {\tt VPSHOCK} model generalizes the {\tt VNEI} one, as it integrates over portions of the plasma with
different ionization ages.  Borkowski, Lyerly \&
Reynolds (2001) showed that NEI models with a single value of
$\tau$ are poor descriptions of the so-called ``Sedov models''
(Hamilton, Sarazin \& Chevalier 1983), designed to model
thermal X-ray emission from SNRs.  Instead, plane-parallel models with
multiple temperatures and $\tau$ are good fits to the Sedov models,
further justifying the preference for the {\tt VPSHOCK+VPSHOCK} model.  We have utilized version 2.0 of the NEI models in {\tt XSPEC}, as used by Z06; the files, with updated ``inner shell processes'' (Z06), were kindly provided to us by the author of the models, K. Borkowski (2007, private communication).

The data and model fits are plotted in Figs. \ref{fig:pn} and \ref{fig:mos}.  Details of the fits to the {\it
EPIC-pn} data as well as the derived
parameters are given in Table \ref{tab:fits}.  For the {\tt
VPSHOCK+VPSHOCK} model, the range of ionization ages, $\tau \equiv n_e
t_{\rm{ion}}$, is bounded by $0 \le \tau \le \tau_u$.  Consistent with
previous studies of SNR 1987A in the X-ray, the upper limit to the
ionization age, $\tau_u$, is {\it higher} for the lower temperature
component.  The computed abundances are given relative to their {\it solar values} (Anders \&
Grevesse [1989], hereafter AG89; Wilms, Allen \& McCray [2000], hereafter
W00; see \S\ref{sect:discussion}), as defined in {\tt XSPEC}.
Galactic foreground absorption is fixed at $6 \times 10^{20}$
cm$^{-2}$, following H06.  To account
for absorption by the LMC, we take the elemental abundances to be 0.5
relative to their solar values, with the exception of helium which
is kept at its solar value.  To get a handle on the systematic errors,
we consider both the AG89 and W00 tables.  For meaningful comparisons, output abundance values must
be normalized to values relative to hydrogen using the respective
abundance table used.  In the said version of {\tt XSPEC}, we note that the
help file for the {\tt VPSHOCK} function erroneously lists it as being hardwired to the AG89
tables, even though it is
superseded by the {\tt abund} command (K. Arnaud 2007, private communication).

The errors  are computed using the
{\tt error} function for $\Delta \chi^2 = 2.706$.  In general, the dependence of the confidence level
on $\Delta \chi^2$ is a non-trivial function (Wall \& Jenkins 2003).
Avni (1976) finds that it depends on the number of parameters that are
estimated simultaneously, i.e., ``interesting parameters'', and not on
the total number of parameters in the fitting function.  For one
interesting parameter, the 90\% confidence interval corresponds to
$\Delta \chi^2 = 2.71$; if more interesting parameters are considered,
the required value of $\Delta \chi^2$ increases.  The $\Delta \chi^2 =
2.706$ level is often quoted as the ``90\%'' confidence interval in {\tt XSPEC}
analyses of SNR 1987A --- we wish to point out that this should be regarded with some care.  Following Z06 and H06, the abundances of He, C, Ar, Ca and Ni are held fixed --- He and C values are taken from Lundqvist \& Fransson (1996, hereafter LF96), while Ar, Ca and Ni ones are from Russell \& Dopita (1992).

The {\tt VPSHOCK+VPSHOCK} model is marginally superior to the {\tt
VNEI+VRAYMOND} one for fitting the {\it EPIC-pn} data --- the reduced
chi-square is $\chi^2_r = 1.21$ versus 1.31.  For reference, we note that the calibration spectra for {\it XMM} are typically fitted with a ``goodness'' of $\chi^2_r \approx 1.5$.  To check for
consistency between the different {\it XMM} instruments, we have taken the {\tt VPSHOCK+VPSHOCK} fit to the
{\it EPIC-pn} data, folded it with the corresponding detector response
and compared it with the {\it EPIC-MOS} and {\it RGS} spectra.  All of
the fit parameters are held fixed in the comparison, but the
normalization is allowed to vary via a constant factor.  This yields
$\chi^2_r = 1.52$ ({\it MOS1}), 1.41 ({\it MOS2}) and 1.56 ({\it RGS}); for the {\tt VNEI+VRAYMOND} model, we get $\chi^2_r = 1.45$, 1.46 and 1.54 instead.
Full-fledged and independent fits --- like for the {\it
EPIC-pn} spectra --- to the {\it
EPIC-MOS} data and their corresponding $\chi^2_r$ values are displayed in
Fig. \ref{fig:mos}.  We note that making more parameters free did not
necessarily improve the fits, implying good cross calibration between
the {\it EPIC-pn} and {\it EPIC-MOS} cameras.

The better counting statistics of the {\it EPIC-pn} data --- and its wider
energy coverage compared to the {\it RGS} --- motivate us to use it
as a template for obtaining the temperatures of the two plasma
components.  Moreover, the higher-temperature component lies beyond
the energy range of the {\it RGS}.  From performing the {\tt VPSHOCK+VPSHOCK} fit to the {\it
EPIC-pn} data, we obtain $kT_{\rm{low}} = 0.4$ keV and $kT_{\rm{high}}
= 3.0$ keV, consistent with the values of $\sim 0.5$ and $\sim 3$ keV
by Z06 (2004 Aug/Sept), who also used a two-shock, NEI model.  By contrast, the {\tt VNEI+VRAYMOND} fit to
the {\it EPIC-pn} data gives $kT_{\rm{low}} = 0.3$ keV and
$kT_{\rm{high}} = 2.4$ keV.  An NEI plasma is at a higher temperature than is reflected by the ionization stages of its elements; therefore, modeling it as a plasma in collisional ionization equilibrium under-estimates its temperature.  We do not expect the temperatures of the plasmas to be substantially lower relative to earlier epochs, as the hot spots on the equatorial ring continue to brighten.  A worthwhile future endeavor will be to
obtain $kT_{\rm{low}}$ and $kT_{\rm{high}}$ for all of the existing
{\it Chandra}, {\it XMM} and {\it Suzaku} observations of SNR 1987A, analyzed
self-consistently using the same {\tt VPSHOCK+VPSHOCK} model with both
AG89 and W00 abundance tables.

\subsection{INTEGRATED FLUX}

By integrating the X-ray spectra, one can obtain fluxes in different
sub-bands.  Park et al. (2004) trisected the spectral range into:
the 0.3 --- 0.8 keV sub-band to represent the O line features (``O
band''); the 0.8 --- 1.2 keV sub-band for the Ne line features (``Ne
band''); and the 1.2 --- 8 keV sub-band for the Mg/Si lines plus any
hard-tail emission features (``H band'').  The soft X-ray band is commonly
regarded to be in the 0.5 to 2 keV range (P05; H06), a historical
convention from the era of {\it ROSAT} (Hasinger,
Aschenbach \& Tr\"{u}mper 1996).  The
``hard band'' is taken to be 3 to 10 keV by
P05.  We integrate the {\it uncorrected} X-ray spectra for various sub-bands and
tabulate them (for 1-$\sigma$ confidence intervals) in Table
\ref{tab:fluxes}.  To correct for Galactic and LMC absorption, one
sets the column densities to zero and then re-computes the fluxes.
Due to the different abundance tables used and column densities inferred, the absorption-corrected luminosities are model-dependent.
Assuming a distance to the SNR 1987A of $d \approx 50$ kpc,
we obtain for $L_{\rm{0.5-2 keV}}$: $1.95 \times 10^{36}$ erg s$^{-1}$
({\tt VPSHOCK+VPSHOCK}, W00 table), $2.32 \times 10^{36}$ erg s$^{-1}$
({\tt VPSHOCK+VPSHOCK}, AG89 table) and $2.28 \times 10^{36}$ erg s$^{-1}$
({\tt VNEI+VRAYMOND}, W00 table).  For $L_{\rm{0.5-10 keV}}$, we get:
$2.17 \times 10^{36}$ erg s$^{-1}$ ({\tt VPSHOCK+VPSHOCK}, W00 table),
$2.54 \times 10^{36}$ erg s$^{-1}$ ({\tt VPSHOCK+VPSHOCK}, AG89
table) and $2.49 \times 10^{36}$ erg s$^{-1}$ ({\tt VNEI+VRAYMOND},
W00 table).  For the 2007 Jan observation of P07, $L_{\rm{0.5-10 keV}}
\approx 2.35 \times 10^{36}$ erg s$^{-1}$.

\subsection{LINE FITTING}

For the {\it RGS} observations, we consider only data points between
0.45 and 1.1 keV, the range over which 11 emission lines were previously
listed by H06 (see their Table 5 and also Z06).  We note that 38 lines were
considered by H06, but only 11 were listed so as to facilitate
comparison with the {\it Chandra} studies.  We rebin the spectra such that each bin has a
minimum of 30 photon counts; we find that if the minimum count is 40,
some of the lines are covered by only 2 to 3 bins.  The {\it RGS}
spectra are modeled using a thermal bremsstrahlung component for
the continuum and a set of Gaussian profiles for the lines.  We first fit
only to the 11 lines of N~{\sc vii}, O~{\sc vii}, O~{\sc viii},
Ne~{\sc ix} and Ne~{\sc x} (Fig. \ref{fig:rgs}).  We
see that several lines are not fitted by such a model
--- we propose that 6 additional lines are now observed: the 725, 727,
739, 812 and 826 eV lines of Fe~{\sc xvii}; and possibly the 873 eV
line of Fe~{\sc xviii}.  (See Table 1 of Behar et al. [2001] and
Fig. 2 of Rasmussen et al. [2001] for details on the expected lines in
SNR spectra within the {\it RGS} energy range.)  The detection of the iron lines
lends support to our derived Fe abundance and its implications (see
\S\ref{sect:discussion}).  

Following H06, we force the widths of all
of the lines to be the same.  The more energetic lines are probably
powered by the faster shocks, and should therefore have larger widths.
A more realistic approach is therefore to model the widths as some
increasing function of the ionization potential of each line.
However, the rather narrow energy range considered here does not
warrant such an approach.  The redshift inferred from the fits is $(1.88 \pm 0.26) \times 10^{-3}$.  The Gaussian width of the
lines is $\sigma_{\rm{fit}} = 0.87 \pm 0.11$ eV --- the full width at
half-maximum (FWHM) value is $2.05 \pm 0.26$ eV, which is narrower than the
$5.3 \pm 1.0$ eV value of H06.  The measured Gaussian width is indicative of the velocities of the shocks responsible for the X-ray emission --- the velocities inferred are $v_s \sim 0.87c/E_{\rm{eV}}$, where $E_{\rm{eV}}$ is the energy of a given line in electron volts.  For a 1 keV line, we have $v_s \sim 300$ km s$^{-1}$, consistent with the {\it Chandra} studies of Z06, who find that the X-ray emission originate from shocks with $300 \lesssim v_s \lesssim 1700$ km s$^{-1}$.

The fluxes we obtain from fitting to these lines are given in Table
\ref{tab:lines}.  In Fig. \ref{fig:lineratios}, we show the relative
increases in flux of the lines (or line complexes) measured in our
study versus those of H06.  Since 2003 May, the line
fluxes have increased by factors $\sim 3$ to 10 --- the 775 eV line of
O~{\sc viii} shows the strongest increase.  On
average, the lines show an increase of $6.0 \pm 0.6$ over $\sim 4$ years.
The 2003 May soft X-ray flux was measured by H06 to be $F_{\rm{0.5-2
keV}} = (8.10 \pm 0.09) \times 10^{-13}$ erg cm$^{-2}$ s$^{-1}$.  Our
2007 Jan value is $F_{\rm{0.5-2
keV}} = (3.34 \pm 0.04) \times 10^{-12}$ erg cm$^{-2}$ s$^{-1}$,
corresponding to an increase of $4.1 \pm 0.1$ in the soft X-ray flux.  Generally, the same shocks that are responsible for the brightening of the optical and soft X-ray hot spots are also powering the lines.  The differences in the flux increases may be an indication that the equivalent widths of the lines are changing, which is evidence for the evolution of the elemental abundances.  Such an investigation is deferred to future studies, when one can obtain smaller error bars for the relative increases of the line fluxes (Fig. \ref{fig:lineratios}).

\section{DISCUSSION}
\label{sect:discussion}

\subsection{INDIVIDUAL ABUNDANCES \& THEIR RATIOS}
\label{subsect:abund}

We compare the derived elemental abundances to those of Z06 and H06
and list them {\it relative to hydrogen} in Table
\ref{tab:abundances}; uncertainties in the AG89 and W00 abundance
tables are not propagated.  We emphasize that care must be taken to
specify the abundance table used, as this may lead to widely differing values of the derived abundances (relative to hydrogen).  In modeling the LMC absorption, Z06 used the elemental abundance table of AG89, while H06 chose the table of W00 because the lower oxygen abundance fitted the K absorption edge in the {\it EPIC} data better.

Fransson et al. (1989) found N/O = 1.6 $\pm$ 0.8, about 12 times
higher\footnote{The AG89 and W00 values for solar
nitrogen-to-oxygen abundance are N/O = 0.132 and 0.110 by number, respectively.} than the AG89 solar value, which they interpret as evidence of
substantial CNO processing.  LF96 found N/O = 1.1 $\pm$ 0.4, while Sonneborn et
al. (1997) found N/O = 1.7 $\pm$ 0.5.  All three of these values were
derived from optical/ultraviolet data.  Next, we turn our attention to
the N/O ratio derived from X-ray studies.  Linearly propagating the
errors listed by Z06, we find that their results yield N/O =
$1.10^{+0.47}_{-0.45}$; they remark that their derived C, N and O
abundance, C+N+O $\approx 1.98 \times 10^{-4}$, is lower by about a
factor of 2 compared to the $3.72 \times 10^{-4}$ value of LF96.  Model A ({\tt VNEI+VRAYMOND}) in H06 yields C+N+O
$\approx 1.67 \times 10^{-4}$ and a rather wide range in the
nitrogen-to-oxygen ratio, N/O = $1.33^{+1.47}_{-0.71}$.  Our {\tt
VPSHOCK+VPSHOCK} fit to the {\it EPIC-pn} data yields C+N+O $\approx 1.29
\times 10^{-4}$ and N/O = $1.17^{+0.37}_{-0.34}$ (using the W00
table); we call this combination of N and O the ``best fit point''.
Note that the C abundance is held fixed at 0.09 relative to solar ($\sim 3 \times 10^{-5}$ relative to hydrogen) for the C+N+O values derived from the X-ray studies.

We next perform a more careful analysis of the N/O ratio.  We first generate a $\chi^2$ map quantifying the
inter-dependence of the fits to the N and O abundance.  Contour lines in the $\chi^2$ map form ``error
ellipses'', which are shown for different $\Delta \chi^2$ values from the
best fit point (Fig. \ref{fig:NO}).  At the $\Delta \chi^2 = 2.706$
level, N/O $= 1.17 \pm 0.20$ for the {\it EPIC-pn} data.  In linearly propagating the errors in the individual abundances, one is in essence adopting the largest possible range of
ratios, which can be visualized as the edges of a rectangle in the
contour map.  Our error analysis improves the uncertainties because it
considers only values of the abundance ratio within the specified
contour.  Within the same confidence interval considered, the
corresponding C+N+O value is from $\sim$ 1.1 to $1.4 \times 10^{-4}$.
We see that the errors in the N/O ratio and the
individual N and O abundances can be constrained at the $\sim 20\%$ level.  We
thus confirm the C+N+O under-abundance noted by the {\it Chandra}
studies of Z06, who suggest a couple of physical reasons for such a
result: the sub-LMC abundance of C+N+O within the progenitor star,
Sanduleak -69$^\circ$202, and/or an extra source of possibly non-thermal,
X-ray continuum.

We perform the same analysis for N/S (Fig. \ref{fig:othercontours}).  Again using the W00 table, linear
propagation of the errors in the abundances obtained from the {\tt VPSHOCK+VPSHOCK}
fit yields  N/S = $4.59^{+1.51}_{-1.36}$, while the error ellipse
analysis gives N/S = $4.59^{+1.59}_{-1.33}$.  The iron, nitrogen and oxygen lines are
predominantly from the lower energy part ($\lesssim 1$ keV) of the spectrum, while the
sulphur lines are situated between $\sim 2.2$ and 3.1 keV.  (The silicon lines are located between
$\sim 1.8$ and 2.2 keV.)  Abundance ratios based on
lines of widely differing energies lead to rounder error ellipses ---
``error circles''.  In such cases, a more thorough error analysis will
not constrain the abundance ratio better, as in the case of
N/S.  This is partially an instrumental effect --- when the energy
resolution is comparable to the spacing of the lines, the line
complexes overlap and are only partially resolved.  There is also the issue of choosing a coordinate system in which the fitting parameters are ``orthogonal''.  Hydrogen and helium have no X-ray lines --- their abundances are derived from the strength of the X-ray continuum.  When the pair of elements considered are situated close to each other in energy, the hydrogen abundance has to first order a linear dependence on the continuum strength.  Thus, the ratio of the considered abundances is tightly constrained as the individual abundances track each other closely.  By contrast, when the pair of abundances considered are located far apart in energy, this linear dependence of hydrogen abundance on the continuum is broken as the dominant uncertainties are in temperature rather than in flux.  Orthogonality is now absent.  To attain orthogonality for such pairs of lines, one has to construct models that directly fit to the abundance ratio considered, an approach which is not explored in this paper.

The error in the abundance ratios increases as one moves a given $\Delta \chi^2$
away from the minimum point.  In Fig. \ref{fig:ratioerror}, we compute
the mean error sustained by the various ratios as a function of $\Delta
\chi^2$.  According to Avni (1976), if three interesting
parameters are considered (see \S\ref{subsect:spectral}), the 90\% confidence interval is situated at $\Delta
\chi^2 = 6.25$.  In this case, the N/O abundance ratio suffers from errors
$\sim 25\%$.

\subsection{THE PROGENITOR OF SNR 1987A: SINGLE-STAR OR BINARY MODEL?}
\label{subsect:metallicity}

A more revealing approach to analyze the elemental abundances is to
normalize the results listed in Table \ref{tab:abundances} by
the ``canonical'' values of the LMC
abundances (Hughes, Hayashi \& Koyama 1998).  These were derived using
a sample of 7 middle-aged SNRs in the LMC (N23, N49, N63A, DEM 71,
N132D, 0453-68.5 and N49B).  This approach was first explored by
A07, who showed that the normalized abundances appear to
cluster in two groups: N, O, Ne, Mg and Fe are slightly more than half
their LMC values, while Si, S and Ni exceed their LMC values.

We generalize the A07 approach by considering both sets of
abundances derived from using the AG89 and W00 tables.  The normalized
abundance, relative to its respective LMC value, is $R_{\rm{87A/LMC}}$; it
is plotted as a function of the elemental mass number in
Fig. \ref{fig:abundratios}.  The error bars for $R_{\rm{87A/LMC}}$ are
computed by linearly propagating the errors listed in Table
\ref{tab:abundances} and in Hughes, Hayashi \& Koyama
(1998).  We caution that additional systematic errors may be present
that are not taken into account.  For example, the derived abundance
for Fe is a sensitive function of temperature and may vary
substantially when small changes are made to $T_{\rm{low}}$
\footnote{The Fe abundance obtained from the {\tt VPSHOCK+VPSHOCK} fit (W00
table) varies by $\sim 16\%$ when $T_{\rm{low}}$ is changed by $\sim 10\%$.}.

We see that the elements O, Ne, Mg and Fe are under-abundant, while Si and S are over-abundant, consistent
with the findings of A07.  With the exception of Fe, there is a
tendency for $R_{\rm{87A/LMC}}$ to increase with larger elemental mass
number, a trend that is independent of the AG89 or W00 tables, though
we note that it is more pronounced with the latter.  The Fe abundance
derived is essentially independent of the AG89 or W00
tables, and is under-abundant by about 70\% relative to the LMC
\footnote{The O abundance relative to H for the AG89 and W00 tables
are the same.}.  The under-abundance of Fe and O was previously noted by Hasinger et al. (2006), who argued for the existence of iron-oxygen ``rust grains''.  The reduced abundance of Fe alone suggests that the iron is locked up in dust grains.  However, Dwek \& Arendt (2007) showed from an analysis of the
infrared-to-X-ray flux ratio --- ${\cal R}_{\rm{IRX}} < 1$ versus the theoretically
expected value of $\sim 10^2$ to $10^3$ --- that the dust in SNR 1987A
is severely depleted compared to standard dust-to-gas mass ratios in
the LMC, suggesting low dust condensation efficiency or dust
destruction in the hot X-ray gas.  In fact, ${\cal R}_{\rm{IRX}}$ was shown to {\it decrease} with time, which is direct evidence for dust destruction.  Our derived plasma temperatures are
consistent with this scenario --- even if dust could form, it
would be destroyed at these temperatures.  

In light of Fig. \ref{fig:abundratios}, the central question to ask is whether the progenitor of SNR 1987A arose from a single star or a binary system?  Sanduleak -69$^\circ$202 was known to be a blue supergiant (BSG) at the time of the supernova explosion, contrary to the expectation that massive stars end their lives as red supergiants (RSGs).  Observations of low-velocity, nitrogen-rich circumstellar material are interpreted as the progenitor star being a RSG until about $\sim 20,000$ years before its death (Fransson et al. 1989).  Such a time scale has in turn been interpreted as the Kelvin-Helmholtz time of the helium core (Woosley et al. 1997).  This BSG-RSG-BSG evolution remains one of the greatest challenges for the single-star model (Woosley et al. 1997; Woosley, Heger \& Weaver 2002), as is the observed system of three rings produced $\sim 20,000$ years before the explosion.  The favored single-star models require a combination of reduced metallicity and ``restricted semi-convection'' (Woosley 1988), the former of which is supported by our derived abundance ratios.  An additional challenge for the single-star model is to reproduce the
over-abundance of Si and S (Fig. \ref{fig:abundratios}), which may require the invoking of some non-standard mixing process (H.-T. Janka 2007, private communication).

Binary solutions to Sanduleak -69$^\circ$202 are sub-divided into
accretion and merger models (see Woosley, Heger \& Weaver [2002] and
references therein).  The binary accretion models allow helium- and
nitrogen-rich material to be added to the progenitor star and require
the disappearance of the mass donor in an earlier supernova event.  In
the binary merger scenario proposed by MP0607, two stars with masses
$\sim 5 M_\sun$ and $\sim 15 M_\sun$ are initially orbiting each other
with a period $\sim 10$ years.  The more massive companion transfers
mass to the less massive star only after the former has completed
helium burning in the core.  A common envelope (Paczy\'{n}ski 1976) is
formed, during which core material from the primary star is dredged up
to the surface.  The merger process takes a few hundred years,
culminating in an initially over-sized RSG, which loses its excess
thermal energy over a few thousand years to become a BSG.  The
spun-up, rapidly rotating BSG produces a fast stellar wind, sweeping
up ejecta associated with the merger, producing the triple ring nebula
we now see in projection.  The nearly axi-symmetric but highly
non-spherical nature of the rings suggests that rotation played a role
in their formation and is consistent with the proposed scenario.  The
beauty of the model lies in the fact that it requires no physically ad
hoc assumptions --- apart from a small kick of $\sim 2$ km s$^{-1}$
given to the ejecta to displace the center of the outer rings from
their symmetry axis --- and makes a number of predictions.  In their
favored model, MP0607 assert that the outer rings are ejected before
the stellar core material is dredged up, while the inner, equatorial
ring --- the site of the observed X-ray emission --- is ejected {\it
afterwards} \footnote{As noted by MP0607, this hypothesis is verifiable/refutable, as the inner ring should exhibit helium enhancement and more CNO processing relative to the outer rings.}.  A relevant consequence of this model is that the dredged-up heavy elements will manifest themselves in the form of X-ray emission lines.  This may explain the trend we see in Fig. \ref{fig:abundratios} --- the challenge for binary merger models is to reproduce the derived $R_{\rm{87A/LMC}}$ values.

Stellar nucleosynthesis can add and {\it not} subtract iron --- in the context of the MP0607 binary merger model, we expect $R_{\rm{87A/LMC}}$(Fe) $\ge 1$.  We are again led to the question: where is the iron?  If the Fe abundance derived is an upper limit on the iron used to form Sanduleak -69$^\circ$202, then it is clearly inconsistent with ``standard'' Fe abundances in the LMC.  An alternative interpretation is that there is
a strong spatial variation in the Fe abundance throughout the LMC,
such that the iron is sub-LMC at the site of SN 1987A and equal to its
LMC abundance elsewhere.  The C+N+O under-abundance suggested in
\S\ref{subsect:abund} supports such a conclusion.  An improvement over using the Hughes, Hayashi \& Koyama (1998) {\it ASCA} abundance values is to re-analyze and expand upon their SNR sample using {\it Chandra} and {\it XMM}.  Existing studies (e.g., Hughes et al. 2006) tend to pick out regions of interest that may include ejecta enrichment; instead, X-ray emission should be extracted from the {\it entire} blast wave, from which average abundances can be inferred.  Future studies will be invaluable towards
resolving these issues.

\scriptsize
K.H. acknowledges the kind hospitality and financial support of: the
Max Planck Institutes for Astrophysics (MPA) and Extraterrestrial
Physics (MPE) during June to October 2007, where he was a visiting
postdoctoral scientist; and the Lorentz Center
(Leiden) during the August 2007 workshop, ``From Massive Stars to
Supernova Remnants''.  He thanks Dick McCray, Sangwook Park, Svet Zhekov, Jack Hughes, John
Raymond, Philipp Podsiadlowski, Rashid Sunyaev, Peter Lundqvist, Claes
Fransson, Dmitrijs Docenko, Thomas Janka, Carlos Badenes, Roger Chevalier, Nathan Smith and Jacco Vink
for engaging and helpful discussions.  Special mentions go out to: Dick and John, who provided a crash course on
non-equilibrium ionization plasmas during a sunny bicycle ride in
Leiden; Svet, who pointed out relevant material on X-ray physics, as well as
for critical comments following his meticulous scrutiny of the manuscript.
The authors are collectively grateful to Kazik Borkowski for providing
the updated {\tt VPSHOCK} model files for use in {\tt XSPEC}.  The
{\it XMM-Newton} project is supported by the {\it Bundesministerium f\"{u}r Wirtschaft und Technologie/Deutsches Zentrum f\"{u}r Luft- und Raumfahrt} (BMWI/DLR, FKZ 50 OX 001) and the Max Planck Society.
\normalsize



\begin{table}
\caption{{\it XMM-Newton} observations of SNR 1987A}
\label{tab:obs}
\begin{center}
\begin{tabular}{llccccc}
\tableline\tableline
\multicolumn{1}{l}{Instrument} &
\multicolumn{1}{c}{Read-out} &
\multicolumn{1}{c}{Filter} &
\multicolumn{1}{c}{Satellite} &
\multicolumn{1}{c}{Date} &
\multicolumn{1}{c}{Time (UT)} &
\multicolumn{1}{c}{Net Exposure} \\
\multicolumn{1}{c}{} &
\multicolumn{1}{c}{Mode} &
\multicolumn{1}{c}{} &
\multicolumn{1}{c}{Revolution} &
\multicolumn{1}{c}{} &
\multicolumn{1}{c}{} &
\multicolumn{1}{c}{(s)} \\
\tableline
{\it EPIC-pn}   & FF, 73 ms & Medium & 1302 & 2007 Jan 17---19 & 19:28---01:18 & 93723\\
{\it EPIC-MOS1} & FF, 2.6 s & Medium & 1302 & 2007 Jan 17---19 & 18:24---01:18 & 109479\\
{\it EPIC-MOS2} & FF, 2.6 s & Medium & 1302 & 2007 Jan 17---19 & 18:24---01:18 & 109560\\
{\it RGS1}      & Spectro   & ---    & 1302 & 2007 Jan 17---19 & 18:23---01:19 & 110528\\
{\it RGS2}      & Spectro   & ---    & 1302 & 2007 Jan 17---19 & 18:23---01:19 & 110475\\
\tableline
\end{tabular}
\end{center}
\end{table}

\begin{table}
\begin{center}
\caption{Comparison of the {\tt VNEI+VRAYMOND} and the {\tt VPSHOCK+VPSHOCK} models}
\label{tab:fits}
\begin{tabular}{lccc}
\tableline\tableline
\multicolumn{1}{c}{Parameter} & \multicolumn{1}{c}{{\tt VNEI+VRAYMOND}} &
\multicolumn{1}{c}{{\tt VPSHOCK+VPSHOCK}} & \multicolumn{1}{c}{{\tt VPSHOCK+VPSHOCK}}\\
\multicolumn{1}{c}{} & \multicolumn{1}{c}{W00$^\spadesuit$} & 
\multicolumn{1}{c}{AG89$^\diamondsuit$} & \multicolumn{1}{c}{W00$^\spadesuit$}\\
\tableline
$\chi^2$ & 820.2032 & 805.6812 & 755.9608\\
$N_{\rm{dof}}^\clubsuit$ & 624 & 623 & 623\\
$\chi^2_r$ & 1.314428 & 1.293228 & 1.213420\\
$kT_{\rm{low}}^\ddagger$ (keV) & 0.320$^{+0.010}_{-0.009}$ & $0.419^{+0.016}_{-0.020}$  & 0.434$^{+0.018}_{-0.008}$\\
$kT_{\rm{high}}^\ddagger$ (keV) & 2.42$^{+0.175}_{-0.176}$ & $2.88^{+0.21}_{-0.15}$ & $2.99^{+0.24}_{-0.22}$\\
Galactic $N_{\rm{H}}$ ($10^{21}$ cm$^{-2}$)$^\dagger$ & 0.6 & 0.6 & 0.6\\
LMC $N_{\rm{H}}$ ($10^{21}$ cm$^{-2}$) & $3.13 \pm 0.14$ &
$2.37 \pm 0.10$ & 2.56$^{+0.07}_{-0.11}$\\
$\tau_u (T_{\rm{low}})$ ($10^{11}$ cm$^{-3}$ s) & --- & $8.63^{+1.60}_{-0.98}$ & $8.33^{+1.02}_{-1.20}$\\
$\tau_u (T_{\rm{high}})$ ($10^{11}$ cm$^{-3}$ s) &
$1.19^{+0.31}_{-0.24}$$^\heartsuit$ & $2.46^{+0.79}_{-0.31}$ & $2.21^{+1.44}_{-0.75}$\\
He$^\dagger$ & 2.57 & 2.57 & 2.57\\
C$^\dagger$ & 0.09 & 0.09 & 0.09\\
N & 0.116$^{+0.172}_{-0.106}$ & $0.303 \pm 0.060$ & 0.536$^{+0.108}_{-0.091}$\\
O &0.0228$^{+0.0074}_{-0.0058}$ & $0.0475^{+0.0045}_{-0.0044}$ & 0.0502$^{+0.0058}_{-0.0060}$\\
Ne & 0.157$^{+0.017}_{-0.015}$ & $0.156^{+0.012}_{-0.007}$ & 0.204$^{+0.017}_{-0.018}$\\
Mg & 0.208$^{+0.021}_{-0.019}$ & $0.116^{+0.007}_{-0.008}$ & 0.168$^{+0.011}_{-0.015}$\\
Si & 0.743$^{+0.068}_{-0.064}$ & $0.262^{+0.025}_{-0.017}$ & 0.479$^{+0.046}_{-0.030}$\\
S & 0.689$^{+0.119}_{-0.111}$ & $0.452^{+0.059}_{-0.077}$ & 0.585$^{+0.075}_{-0.074}$\\
Ar$^\dagger$ & 0.54 & 0.54 & 0.54\\
Ca$^\dagger$ & 0.34 & 0.34 & 0.34\\
Fe & 0.0397$^{+0.0038}_{-0.0033}$ & $0.0630^{+0.0041}_{-0.0018}$ & 0.0895$^{+0.0050}_{-0.0060}$\\
Ni$^\dagger$ & 0.62 & 0.62 & 0.62\\
\tableline
\end{tabular}
\end{center}
\scriptsize
Note: Abundances of elements are listed relative to solar.\\
$\dagger$ Parameter is held fixed (see text).\\
$\ddagger$ In the {\tt VNEI+VRAYMOND} model, the low and high temperature components are from the {\tt VRAYMOND} and {\tt VNEI} fits, respectively.\\
$\clubsuit$ Number of degrees of freedom in fit.\\
$\diamondsuit$ Abundance table in {\tt XSPEC} from Anders \& Grevesse (1989).\\
$\spadesuit$ Abundance table in {\tt XSPEC} from Wilms, Allen \& McCray (2000).\\
$\heartsuit$ {\tt VNEI} model considers only one value of $\tau = \tau_u$.\\
\normalsize
\end{table}

\begin{table}
\begin{center}
\caption{Elemental abundances from different studies (relative to hydrogen)}
\label{tab:abundances}
\begin{tabular}{lcccc}
\tableline\tableline
\multicolumn{1}{c}{Element} & \multicolumn{1}{c}{Z06$^\diamondsuit$} &
\multicolumn{1}{c}{H06$^\spadesuit$} & \multicolumn{1}{c}{This Study$^\diamondsuit$} &
\multicolumn{1}{c}{This Study$^\spadesuit$}\\
\multicolumn{1}{c}{} & \multicolumn{1}{c}{{\tt VPSHOCK+VPSHOCK}} & 
\multicolumn{1}{c}{{\tt VNEI+VRAYMOND}} & 
\multicolumn{1}{c}{{\tt VPSHOCK+VPSHOCK}} & 
\multicolumn{1}{c}{{\tt VPSHOCK+VPSHOCK}}\\
\tableline
N & $\left(8.64^{+2.58}_{-1.91}\right)_{-5}$ &
$\left(7.47^{+4.85}_{-2.52}\right)_{-5}$ & $\left(3.40 \pm 0.67\right)_{-5}$ & $\left(5.00^{+1.01}_{-0.85}\right)_{-5}$\\
O & $\left(7.83^{+1.02}_{-1.48}\right)_{-5}$ & $\left(5.62^{+2.55}_{-1.11}\right)_{-5}$ & $\left(4.04^{+0.38}_{-0.37}\right)_{-5}$ & $\left(4.27^{+0.49}_{-0.51}\right)_{-5}$\\
Ne & $\left(3.57^{+0.62}_{-0.49}\right)_{-5}$ & $\left(2.95^{+0.86}_{-0.49}\right)_{-5}$ & $\left(1.92^{+0.15}_{-0.09}\right)_{-5}$ & $\left(2.51^{+0.21}_{-0.22}\right)_{-5}$\\
Mg & $\left(9.12 \pm 1.52\right)_{-6}$ & $\left(1.05^{+0.31}_{-0.23}\right)_{-5}$ & $\left(4.41^{+0.27}_{-0.30}\right)_{-6}$ & $\left(6.54^{+0.43}_{-0.58}\right)_{-6}$\\
Si & $\left(9.93^{+1.42}_{-2.13}\right)_{-6}$ & $\left(2.34^{+0.57}_{-0.43}\right)_{-5}$ & $\left(9.30^{+0.89}_{-0.60}\right)_{-6}$ & $\left(1.70^{+0.16}_{-0.11}\right)_{-5}$\\
S & $\left(7.30^{+2.43}_{-2.11}\right)_{-6}$ &
$\left(5.96^{+3.72}_{-3.35}\right)_{-6}$ &
$\left(7.33^{+0.96}_{-1.25}\right)_{-6}$ & $\left(1.09 \pm 0.14\right)_{-5}$\\
Fe & $\left(7.48^{+0.47}_{-0.94}\right)_{-6}$ & $\left(1.33^{+0.41}_{-0.28}\right)_{-6}$ & $\left(2.95^{+0.19}_{-0.08}\right)_{-6}$ & $\left(2.83^{+0.16}_{-0.19}\right)_{-6}$\\
\tableline
\end{tabular}
\end{center}
\scriptsize
Note: $A_b$ is shorthand notation for $A \times 10^b$.\\
$\diamondsuit$ Abundance table in {\tt XSPEC} from Anders \& Grevesse (1989).\\
$\spadesuit$ Abundance table in {\tt XSPEC} from Wilms, Allen \& McCray (2000).
\normalsize
\end{table}

\begin{table}
\begin{center}
\caption{Absorption-uncorrected {\it EPIC-pn} fluxes in various sub-bands}
\label{tab:fluxes}
\begin{tabular}{lcc}
\tableline\tableline
\multicolumn{1}{c}{Sub-band} & \multicolumn{1}{c}{photons cm$^{-2}$
s$^{-1}$} & \multicolumn{1}{c}{erg cm$^{-2}$ s$^{-1}$}\\
\tableline
0.2 --- 0.8 keV & $\left(1.16 \pm 0.01 \right)_{-3}$ &
$\left(1.14 \pm 0.01\right)_{-12}$\\
0.8 --- 1.2 keV & $\left(1.01^{+0.02}_{-0.01}\right)_{-3}$ & $\left(1.54^{+0.03}_{-0.06}\right)_{-12}$\\
1.2 --- 8 keV & $\left(4.93^{+0.11}_{-0.15}\right)_{-4}$ & $\left(1.53^{+0.04}_{-0.05}\right)_{-12}$\\
0.5 --- 2 keV & $\left(2.28 \pm 0.02\right)_{-3}$ & $\left(3.34 \pm 0.04\right)_{-12}$\\
3 --- 10 keV & $\left(5.31^{+0.16}_{-0.34}\right)_{-5}$ & $\left(3.83^{+0.11}_{-0.23}\right)_{-13}$\\
0.5 --- 10 keV & $\left(2.42^{+0.02}_{-0.03}\right)_{-3}$ &
$\left(4.05 \pm 0.07\right)_{-12}$\\
0.2 --- 10 keV & $\left(2.67^{+0.03}_{-0.02}\right)_{-3}$ & $\left(4.24^{+0.06}_{-0.05}\right)_{-12}$\\
\tableline
\end{tabular}
\end{center}
\scriptsize
Note: 1-$\sigma$ (68\%) confidence intervals; $A_b$ is shorthand notation for $A \times 10^b$.\\
\normalsize
\end{table}

\begin{table}
\begin{center}
\caption{Line fluxes from {\it RGS} observations}
\label{tab:lines}
\begin{tabular}{lcc}
\tableline\tableline
\multicolumn{1}{c}{Line} & \multicolumn{1}{c}{Energy} &
\multicolumn{1}{c}{Flux}\\
\multicolumn{1}{c}{} & \multicolumn{1}{c}{(eV)} &
\multicolumn{1}{c}{($10^{-5}$ photons cm$^{-2}$ s$^{-1}$)}\\
\tableline
N~{\sc vii} & 500 & $18.67^{+1.45}_{-1.42}$\\
O~{\sc vii} & 561 & $4.94^{+1.19}_{-1.73}$\\
O~{\sc vii} & 569 & $1.05^{+2.12}_{-1.05}$\\
O~{\sc vii} & 574 & $8.92^{+1.14}_{-2.37}$\\
O~{\sc viii} & 654 & $25.18^{+2.13}_{-1.14}$\\
O~{\sc vii} & 666 & $3.21^{+0.65}_{-1.37}$\\
Fe~{\sc xvii} & 725 & $3.83^{+0.74}_{-1.83}$\\
Fe~{\sc xvii} & 727 & $12.98^{+1.17}_{-2.05}$\\
Fe~{\sc xvii} & 739 & $7.12^{+0.63}_{-1.46}$\\
O~{\sc viii} & 775 & $9.40^{+0.84}_{-1.12}$\\
Fe~{\sc xvii} & 812 & $5.70^{+2.31}_{-0.47}$\\
Fe~{\sc xvii} & 826 & $16.00^{+2.43}_{-0.61}$\\
Fe~{\sc xviii} & 873 & $6.00^{+0.77}_{-1.10}$\\
Ne~{\sc ix} & 905 & $8.96^{+1.16}_{-1.63}$\\
Ne~{\sc ix} & 915 & $1.81^{+2.58}_{-0.82}$\\
Ne~{\sc ix} & 922 & $18.87^{+1.17}_{-2.86}$\\
Ne~{\sc x} & 1022 & $17.75^{+1.35}_{-2.37}$\\
\tableline
O~{\sc vii} & 561+569+574 & $14.91^{+4.45}_{-5.15}$\\
Ne~{\sc ix} & 905+915+922 & $29.64^{+4.91}_{-5.31}$\\
\tableline
\end{tabular}
\end{center}
\scriptsize
Note: $\Delta \chi^2 = 2.706$ confidence intervals.\\
\normalsize
\end{table}


\begin{figure}
\begin{center}
\end{center}
\caption{Schematic representation of the physical configuration in SNR
1987A.  Courtesy of Richard McCray (JILA, University of Colorado) and the {\it Chandra} press release team.  {\it Figure has been omitted due to large file size; please refer to the electronic version of the Astrophysical Journal for this figure.}}
\label{fig:87a}
\end{figure}

\begin{figure}
\begin{center}
\end{center}
\caption{{\it XMM-Newton EPIC} image of the region around SNR 1987A. The red-green-blue (RGB) color image is composed of images from three energy bands (0.2---1.0, 1.0---2.0 and 2.0---4.5 keV) and from all three of the {\it EPIC} instruments ({\it PN, MOS1 and MOS2}).  The individual images are exposure-corrected and out-of-time event-subtracted (for {\it EPIC-pn}).  The source and background extraction regions used for the spectral analysis are denoted by ``1987A'' and ``bg'', respectively.  The blue, arc-like structures are due to photons from the bright X-ray binary LMC X-1 (located outside the field-of-view), which are singly-reflected on the mirror shells.  {\it Figure has been omitted due to large file size; please refer to the electronic version of the Astrophysical Journal for this figure.}}
\label{fig:xmm87a}
\end{figure}

\begin{figure}
\begin{center}
\includegraphics[width=5.5in]{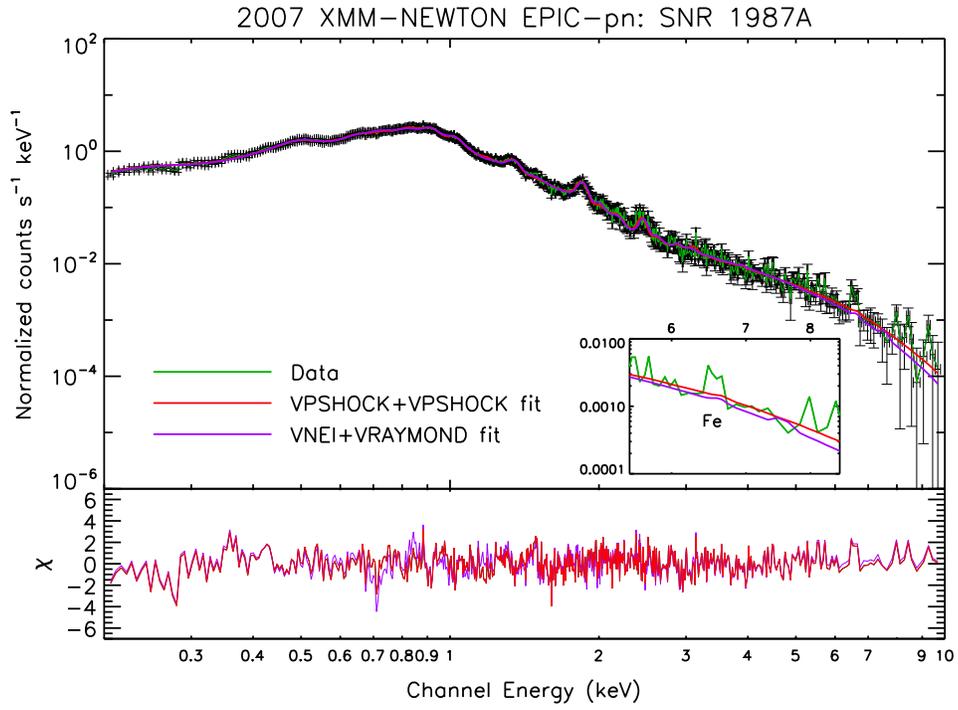}
\end{center}
\caption{{\it XMM EPIC-pn} data, fitted by the {\tt VNEI+VRAYMOND} and {\tt VPSHOCK+VPSHOCK} models in {\tt XSPEC}.  The insert zooms in on a possible detection of the Fe K$\alpha$ line.  Details of the model fits, as well as the derived parameter values, are given in Table \ref{tab:fits}.}
\label{fig:pn}
\end{figure}

\begin{figure}
\begin{center}
\includegraphics[width=5in]{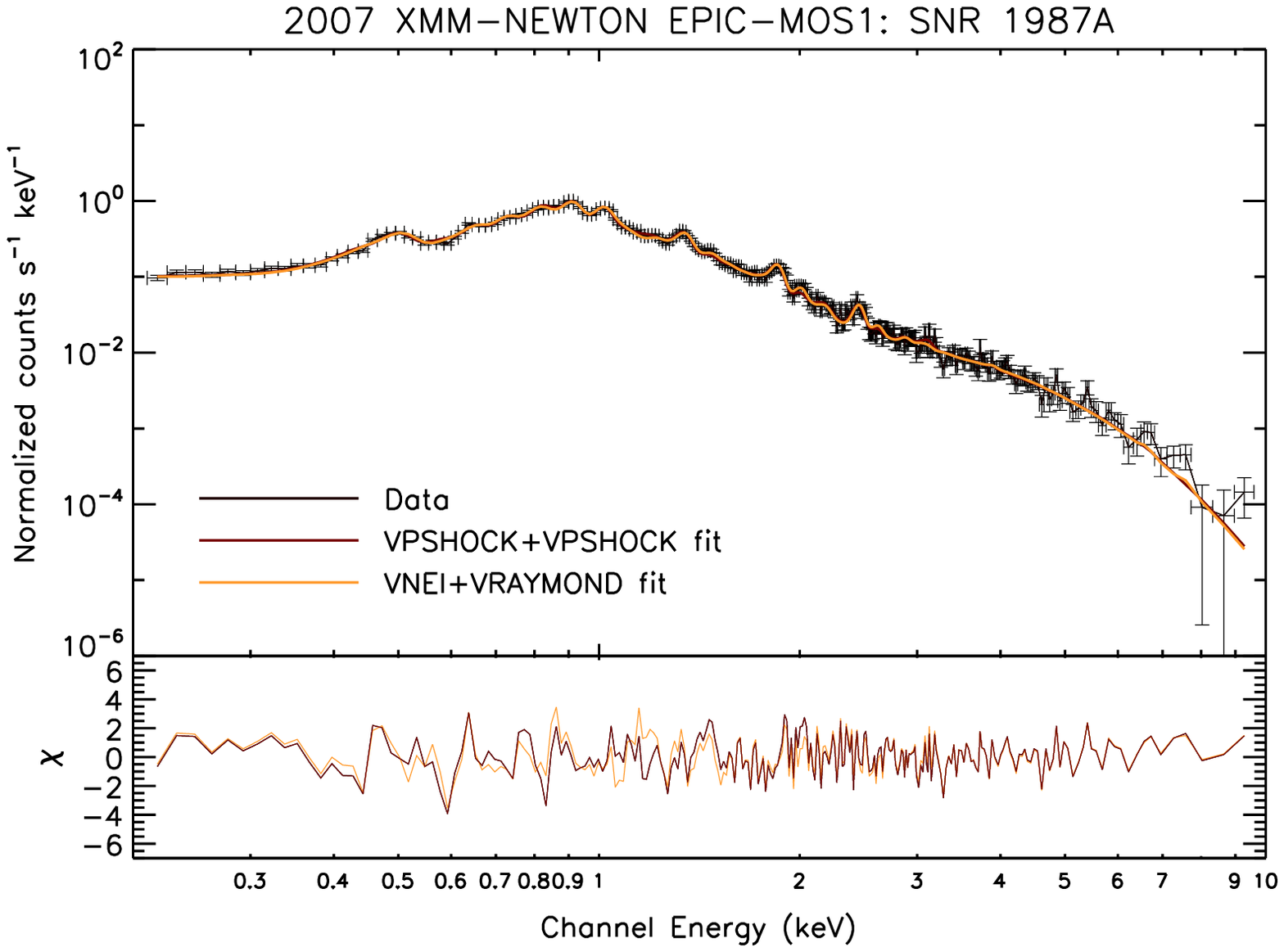}
\includegraphics[width=5in]{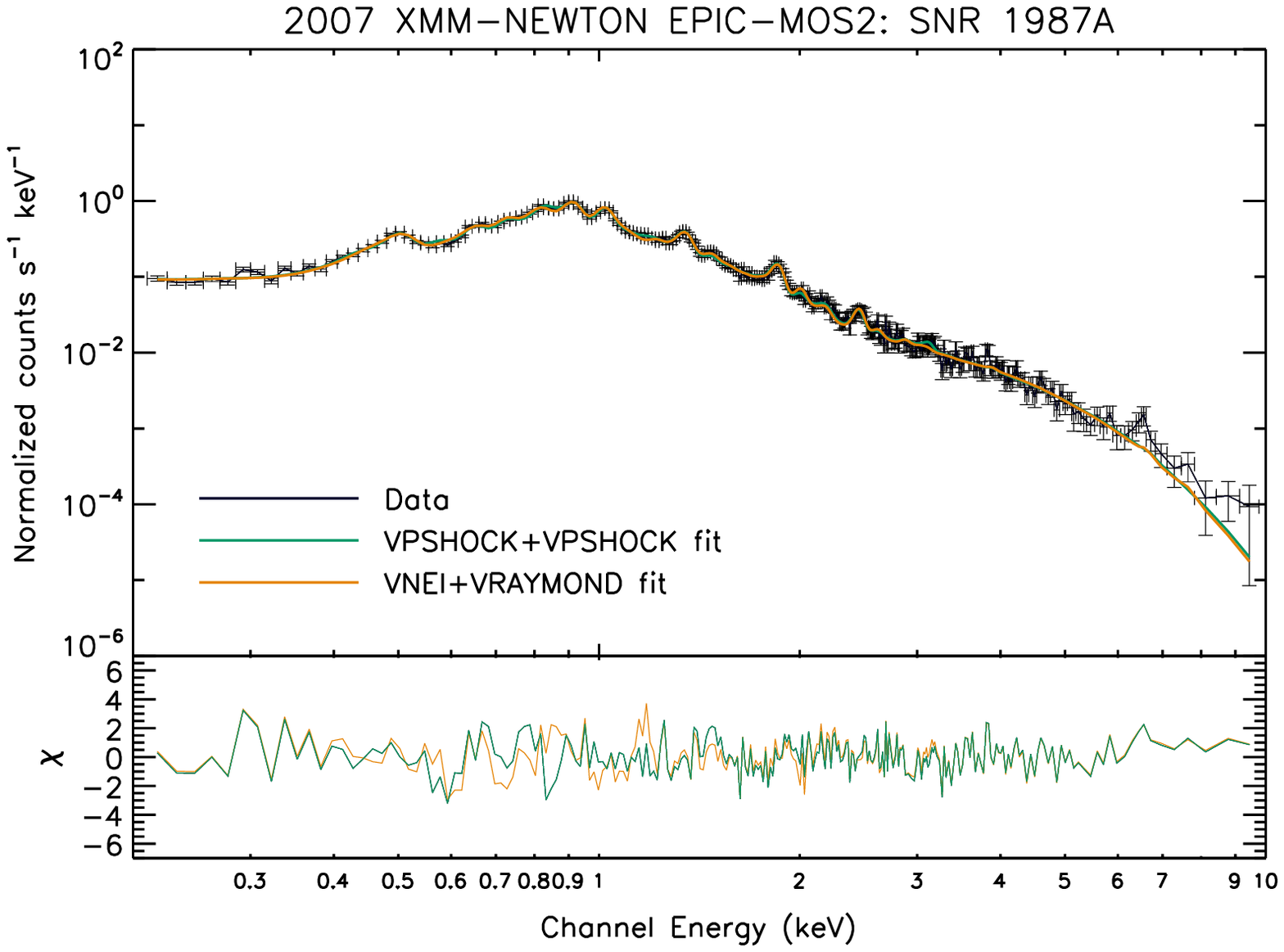}
\end{center}
\caption{{\it XMM EPIC-MOS1} (top) and {\it EPIC-MOS2} (bottom) data, fitted by the {\tt VPSHOCK+VPSHOCK} model.  For the {\tt VPSHOCK+VPSHOCK} model, the reduced chi-square values are $\chi^2_r = 1.53$ ({\it MOS1}) and 1.40 ({\it MOS2}).   For the {\tt VNEI+VRAYMOND} model, the reduced chi-square values are $\chi^2_r = 1.45$ ({\it MOS1}) and 1.46 ({\it MOS2}).}
\label{fig:mos}
\end{figure}

\begin{figure}
\begin{center}
\includegraphics[width=5in]{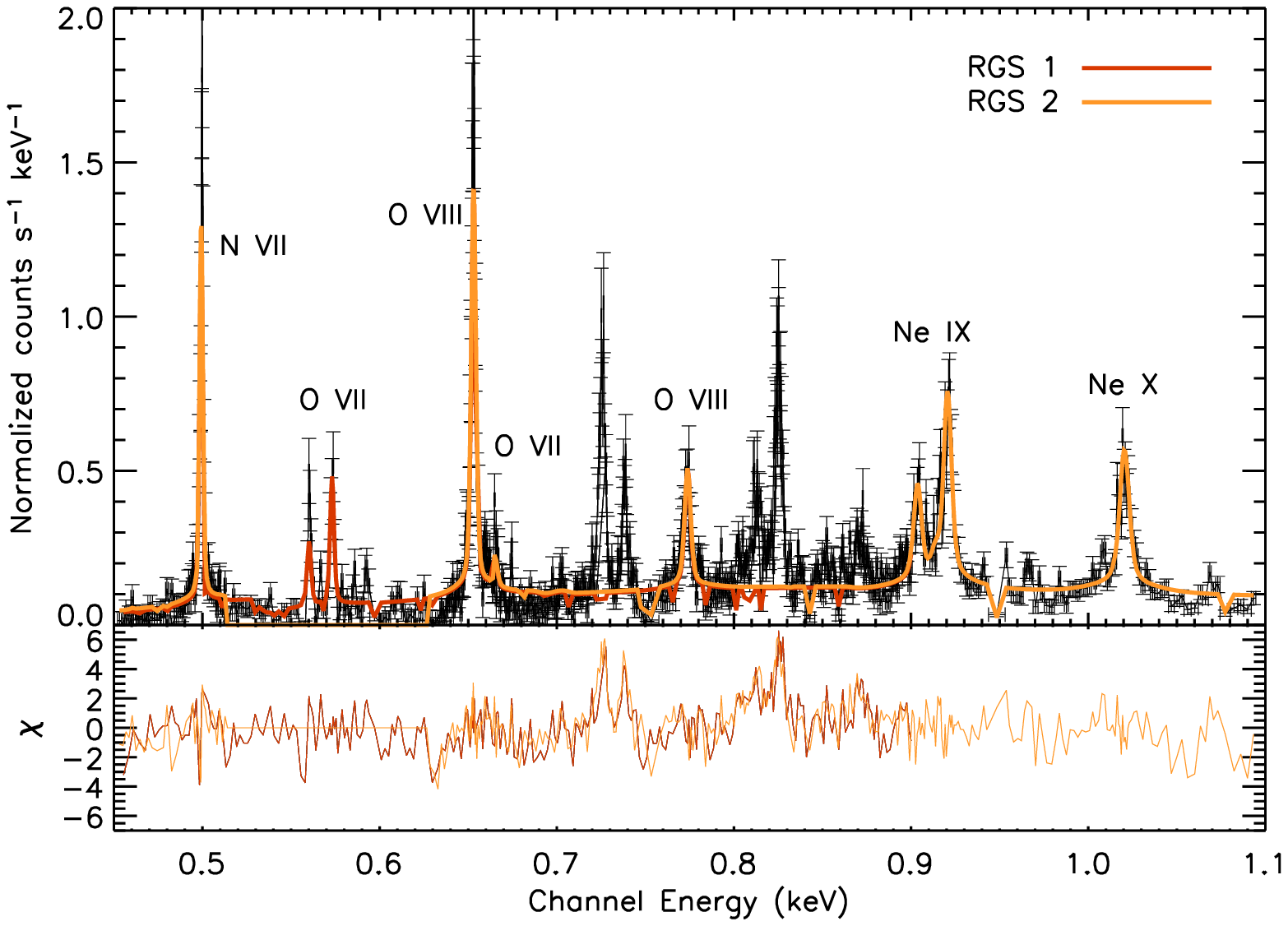}
\includegraphics[width=5in]{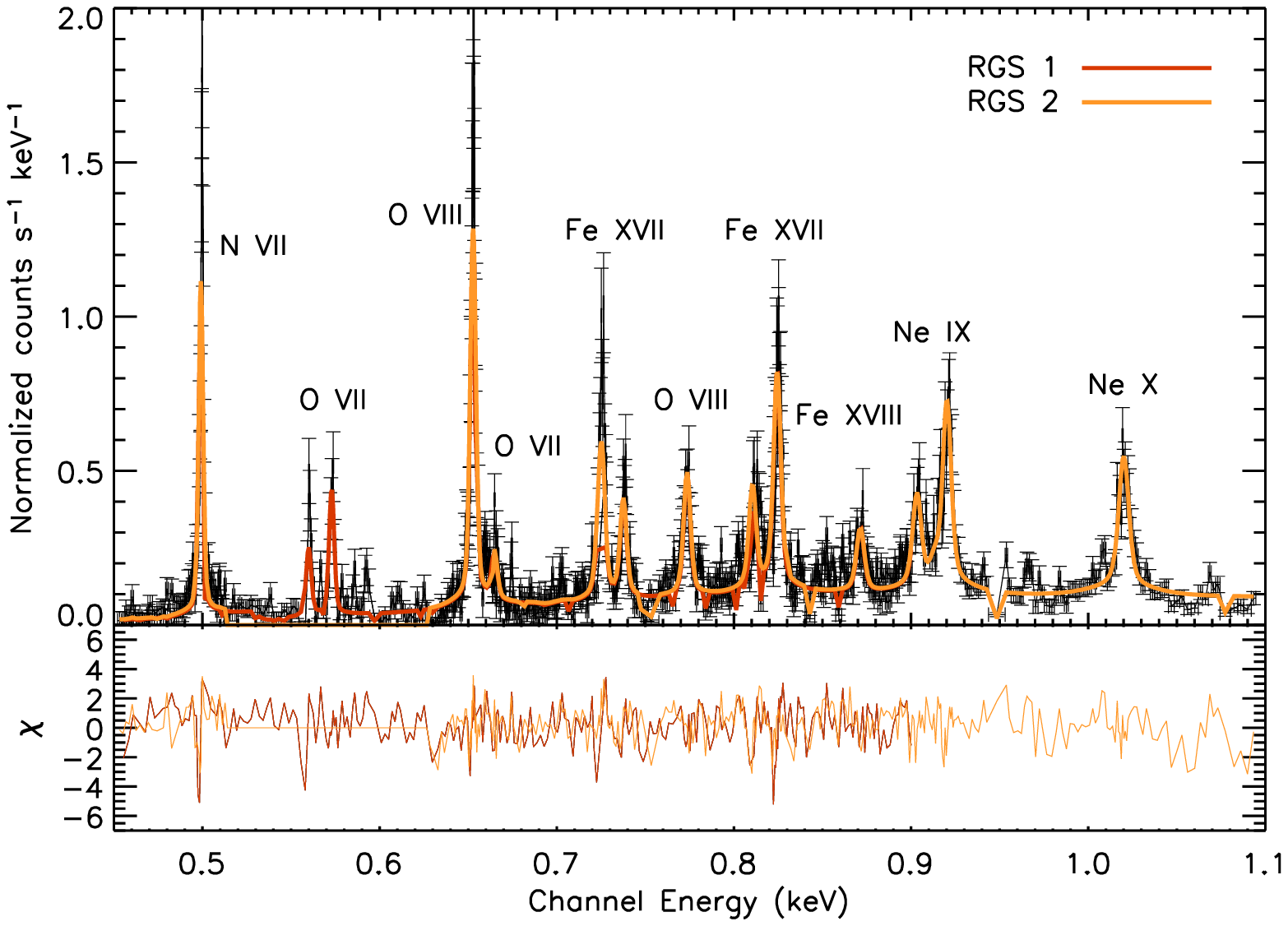}
\end{center}
\caption{{\it XMM RGS} data considering 11 (top; $\chi^2_r = 3.37$ for
$N_{\rm{dof}}=652$) and 17 (bottom; $\chi^2_r = 1.92$ for
$N_{\rm{dof}}=645$) Gaussian fits to the lines, respectively (see text).}
\label{fig:rgs}
\end{figure}

\begin{figure}
\begin{center}
\includegraphics[width=5.5in]{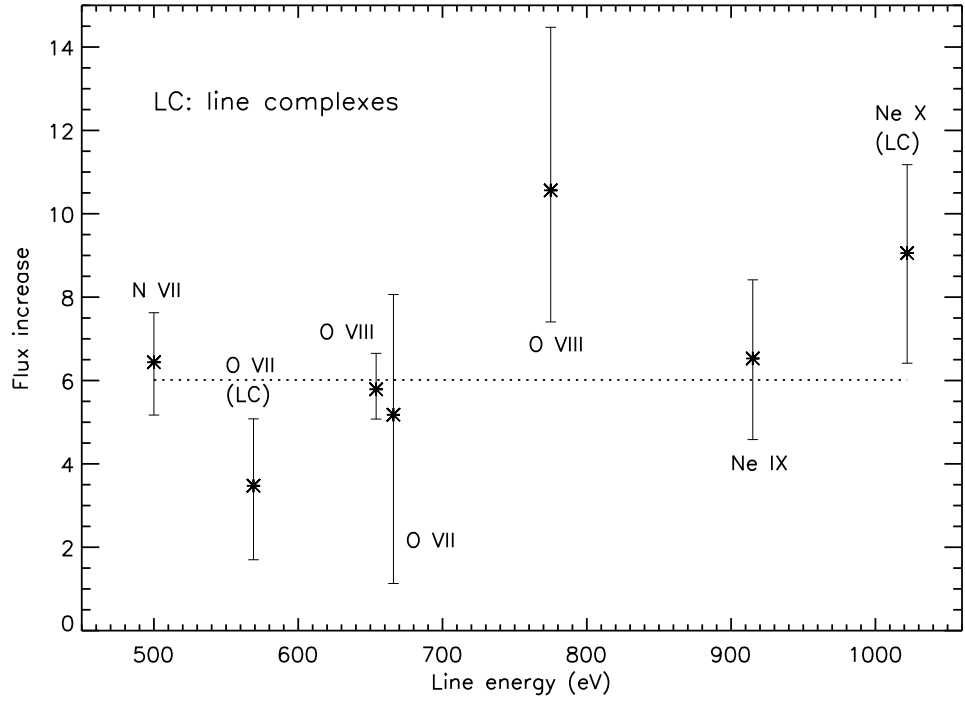}
\end{center}
\caption{Flux increase in emission lines from {\it XMM RGS} data
between 2003 May (Haberl et al. 2006) and this study (2007 Jan).  The line complexes
(denoted ``LC'') represent the summed fluxes from three lines.  See Table
\ref{tab:lines} and the text for details of the lines.  The
horizontal, dotted line shows a fit for the average increase in the
line fluxes: $6.0 \pm 0.6$ $\left(\chi^2_r = 0.98 \right)$.}
\label{fig:lineratios}
\end{figure}

\begin{figure}
\begin{center}
\includegraphics[width=5in]{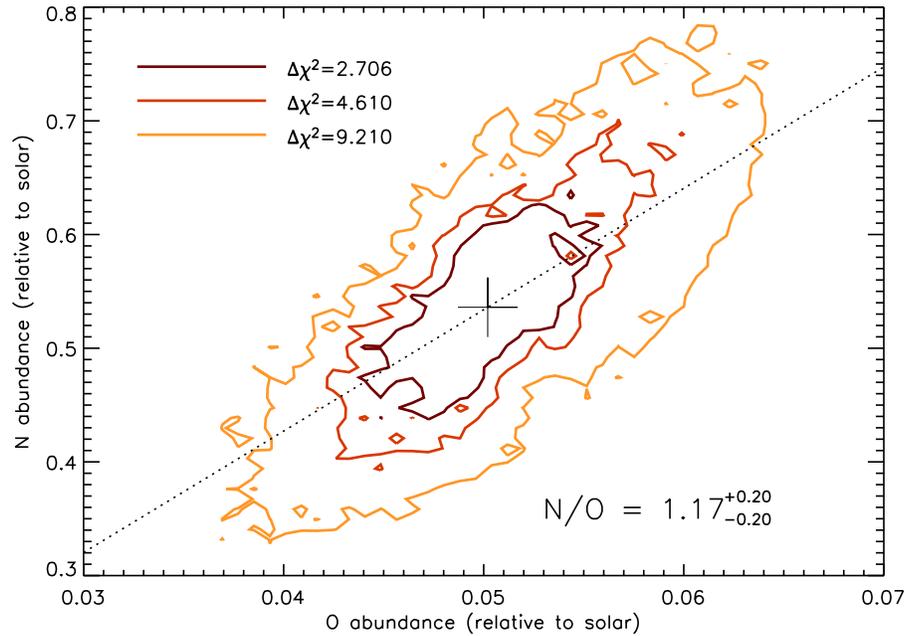}
\end{center}
\caption{Contour map of N and O for various $\Delta \chi^2$ values from the
best fit point, marked by the large cross.  Values of N and O residing along the dotted line have the same N/O value as that for the best fit point.}
\label{fig:NO}
\end{figure}

\begin{figure}
\begin{center}
\includegraphics[width=5in]{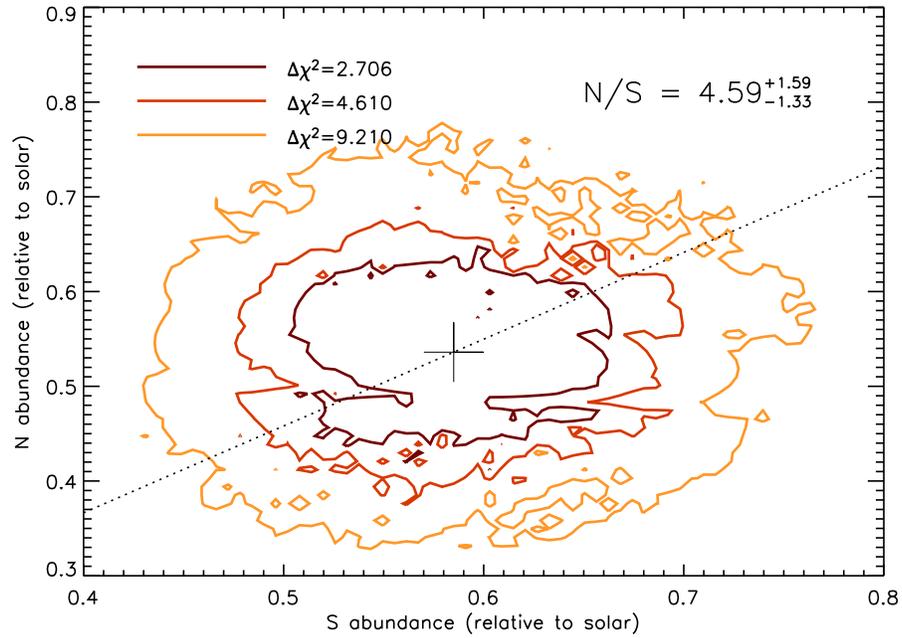}
\end{center}
\caption{Same as Fig. \ref{fig:NO}, but for N/S.}
\label{fig:othercontours}
\end{figure}

\begin{figure}
\begin{center}
\includegraphics[width=5.5in]{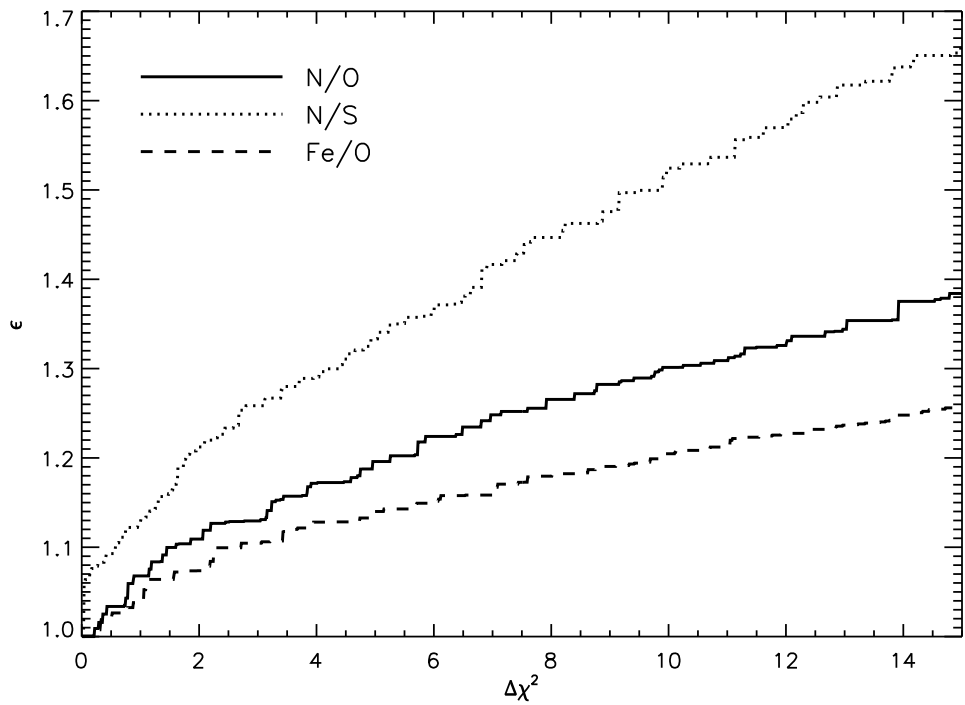}
\end{center}
\caption{Mean error sustained by various abundance ratios as one
moves $\Delta \chi^2$ away from the best fit point in the respective $\chi^2$ maps.}
\label{fig:ratioerror}
\end{figure}

\begin{figure}
\begin{center}
\includegraphics[width=5.5in]{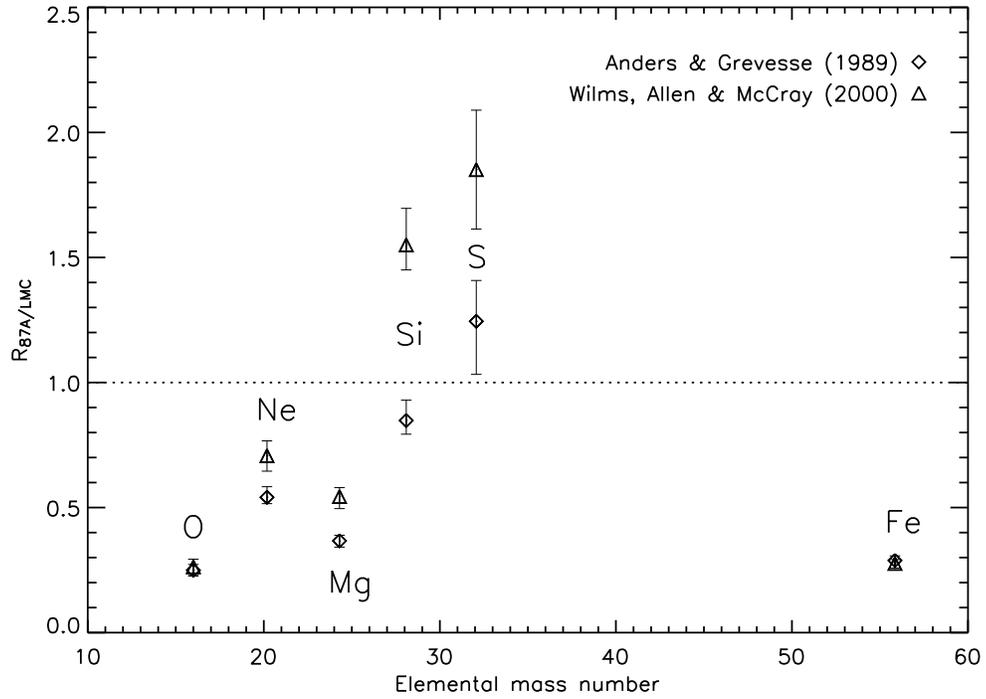}
\end{center}
\caption{Abundance ratios of elements found by our study of SNR 1987A to the LMC quantities derived by Hughes, Hayashi \& Koyama (1998).  In our {\tt XSPEC} analysis, we examine abundances obtained using the tables of Anders \& Grevesse (1989) and Wilms, Allen \& McCray (2000).}
\label{fig:abundratios}
\end{figure}

\end{document}